\documentclass[prb,twocolumn,superscriptaddress,showpacs,amsmath,amssymb]{revtex4}

\usepackage{bm}

\usepackage{graphicx}
\usepackage{latexsym}
\usepackage{amsmath}
\usepackage{amssymb}
\usepackage{amsfonts}
\usepackage{color}
\usepackage{bm}
\usepackage{verbatim}

\begin{document}
\title{Nucleation of domain walls by $Z_2$ symmetry
breaking transition in $p_x+ip_y$ superconductors.}

\author{Vasily Vadimov}
\affiliation{Institute for Physics of Microstructures RAS, 603950
Nizhny Novgorod, Russia.}
\author{Mihail Silaev}
\affiliation{Institute for Physics of Microstructures RAS, 603950
Nizhny Novgorod, Russia.}

\date{\today}

\begin{abstract}
We show that time reversal symmetry breaking $p_x+ip_y$ wave
superconductors undergo several phase transitions subjected to
external magnetic field or supercurrent. In such system the
discrete $Z_2$ symmetry can recover before the complete
destruction of the order parameter. The topological defects
associated with $Z_2$ symmetry - domain walls can be created in a
controllable way by magnetic field or current sweep according to
the Kibble-Zurek scenario. Such domain wall generation can take
place in exotic superconductors like $Sr_2RuO_4$ and some heavy
fermion compounds.
\end{abstract}


\maketitle

Topological defect formation in the systems which undergo
non-equilibrium phase transitions has become a subject to
interdisciplinary research between high energy and condensed
matter physics\cite{VilenkinBook,VolovikBook,KibbleReview}.
  Commonly accepted cosmological model suggests that
cosmic strings can form according to the Kibble-Zurek (KZ)
scenario through the nonequilibrium phase transition in expanding
Universe\cite{Kibble,Zurek}.
 The KZ mechanism was confirmed in experiments with quantized
 vortices in superfluid $^4$He\cite{He4} and $^3$He\cite{EltsovReview,He3}
 which can be produced by rapid quench or
 pressure sweep driving the system through the second order $U(1)$
 symmetry breaking phase transition\cite{KibbleReview}.

The physics of domain walls (DWs) is less studied and remains a
large enigma both in cosmology and condensed matter
systems\cite{Zeldovich,Kibble}. Indeed the observational
constrains require to accept the fact that DWs have disappeared at
the early history of the Universe. A plausible explanation
involves assumptions of the initial baryon asymmetry or  time
inversion symmetry violation which finally totally removes the
domains of one kind\cite{Zeldovich}. However these speculations
remains yet unconfirmed which make theorists to rule out the
models with discrete symmetry breaking since the mechanism of DWs
disappearance remains a mystery.\cite{Kibble}

One of the few known condensed matter systems
 which allows studying quench
induced formation of cosmiclike DWs
 is superfluid $^3$He \cite{VolovikSalomaaDW}.
Experimentally DW generation was detected during the cooling into
A-phase\cite{He3Soliton}. However with rapid temperature sweep one
can hardly fine tune the parameters in order to produce
exclusively DWs without producing vortices and composite
defects\cite{He3Composite}. Moreover in real system quench is
always spatially inhomogoneous which provides important
modifications to the physics of defect formation
\cite{VolovikKibble,ZurekNonHom,KopninThuneberg,KopninAranson}. In
this Letter we
 propose a unique selective mechanism of DWs formation
 during spatially homogeneous phase transition in exotic superconductors with chiral $p_x+ip_y$ pairing symmetry.

  This mechanism is likely to be tested in recently discovered
  layered-perovskite superconductor $Sr_2RuO_4$
  \cite{MaenoNature,MaenoRMP}.
  According to a number of experimental evidences\cite{Exp1,Exp2,MaenoRMP}
  $Sr_2RuO_4$ is assumed to be a chiral $p_x+ip_y$ wave
  superconductor with Cooper pairs having an effective
  internal orbital momentum projection on the
  crystal anisotropy axis $L_z=\pm 1$.
 Such
 superconducting state has a broken time reversal symmetry (TRS) so the superconducting phase transition is determined by
  the spontaneous $U(1)\times Z_2$ symmetry violation. Recently such state was suggested
to appear also in multiband superconductors \cite{EgorZ2}.

The two different TRS breaking vacuum states can be separated by
DWs which are known to support spontaneous supercurrent generating
magnetic fields \cite{MagneticFlux}. However high resolution
scanning SQUID microscopy experiments
 detected no stray fields which should be generated by DWs above the surface
 of superconducting $Sr_2RuO_4$\cite{Kallin}. Moreover polar Kerr effect measurements\cite{Kerr}
 also did not reveal chiral domains.
 Thus up to now no direct observation of DWs in $Sr_2RuO_4$ was
 obtained although phase-sensitive Josephson spectroscopy
  experiments\cite{Josephson} revealed some evidences of dynamical domain structure.
  This enigma of DWs
  stimulated further theoretical research. It has been
  suggested that in some cases the DW generates only very weak stray field\cite{Sonin}. The stray fields
  suppression can result also from the multiband superconductivity
  \cite{StanfordTheory} which on the other hand can stimulate the
  proposed unconventional mixed state with vortex coalescence in $Sr_2RuO_4$
  \cite{BabaevAgterberg}.

  In addition to the above mentioned hypotheses the possibility still
   remains that DWs disappear at some stage of the
  superconducting transition in $Sr_2RuO_4$. Therefore the proposed method
 to create in controllable way an arbitrary initial concentration of DWs in
  $Sr_2RuO_4$ can prompt experimental identification of this defects which
  has been recently one of the most intriguing problems in the field of low
  temperature physics. Moreover it can shed a new light on the
  fate of cosmic DWs during the early history of the
  Universe.

  To describe DWs
  separating different $L_z=\pm 1$ vacuum states
 we use Ginzburg-Landau (GL) model of superconducting state in $Sr_2RuO_4$.
 This material belongs to the tetragonal crystallographic symmetry group $D_{4h}$ and has strong
crystal anisotropy which keeps both spin and orbital momentum of
Cooper pairs parallel to the $c$ axis\cite{MaenoRMP}. The
coordinate system is chosen so that the crystal anisotropy axis is
${\bf c}\parallel {\bf z}$. Then $p_x+ip_y$ state corresponds to
the two-dimensional representation $\Gamma_5^-=({k_x {\bf z}, k_y
{\bf z}})$ and the order parameter is described by a complex
two-dimensional vector
$\eta=(\eta_x,\eta_y)$\cite{SigristUeda,MaenoRMP,Joynt}. Thus
introducing chiral order parameter components $\eta_\pm=\eta_x \pm
i \eta_y$ we consider a GL free energy density in usual
dimensionless units:
\begin{eqnarray}
    \nonumber
    f = &-|{\eta_+}|^2 - |{\eta_-}|^2 +(|{\eta_+}|^4 +
        |{\eta_-}|^4)/2 + 2 |\eta_+\eta_-|^2
        +\\ \label{eq:free_energy_density}
        &[\nu_1  (\eta_- \eta^*_+)^2  + c.c.]/2 +
        \left|{\mathbf D \eta_+}\right|^2 + \left|{\mathbf D \eta_-}\right|^2 +
        \\ \nonumber
        &[{(D_- \eta_+)}^*\left(D_+ \eta_-\right) +
         \nu_2 {\left(D_+ \eta_+\right)}^* \left(D_-
     \eta_-\right) +c.c.]/2
\end{eqnarray}
where $\mathbf D = -i \nabla / \kappa - \mathbf A$, $D_\pm = D_x
\pm i D_y$, ${\bf A}$ is vector potential and $\kappa$ is GL
parameter. Coefficients $\nu_{1,2}$ determine the anisotropy in
$xy$ plane induced by tetragonal distortions. In case if
$\nu_1=\nu_2$ the free energy (\ref{eq:free_energy_density}) was
obtained from weak coupling microscopic theory\cite{AgterbergWC}.

 \begin{figure}[!h]
 \centerline{\includegraphics[width=1.0\linewidth]{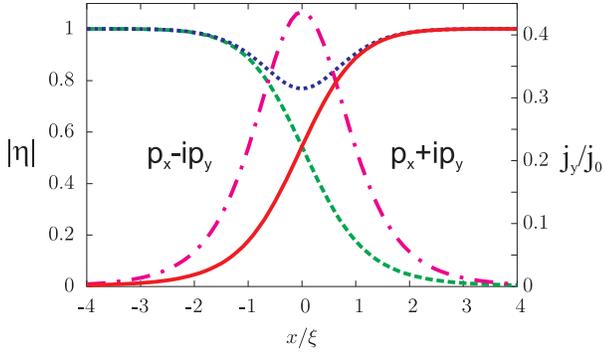}}
 \caption{\label{Fig:Wall} Domain wall structure in $p_x+ip_y$ superconductor described by GL model (\ref{eq:free_energy_density}).
  The DW plane is $yz$. We choose anisotropy parameters $\nu_1=\nu_2=0.1$.
  By solid and dashed lines the distributions $\eta_+(x)$ and $\eta_-(x)$ are shown.
  By dash-dotted line the longitudinal superfluid current density $j_y(x)$ is shown normalized to
  $j_0=(c/4\pi)H_{cm}/\sqrt{2}\xi$.
  The overall  order parameter magnitude
  $\sqrt{|\eta_+|^2+|\eta_-|^2}$ is shown by dotted line.}
 \end{figure}

The GL model (\ref{eq:free_energy_density}) yields two degenerate
ground states $(\eta_+,\eta_-)=(0,1)$ and $(1,0)$. Here we
implement numerical minimization of the GL energy
(\ref{eq:free_energy_density}) choosing the $x$ axis perpendicular
to the DW plane.
 In Fig.\ref{Fig:Wall} we plot the calculated order parameters and equilibrium density of
  supercurrent which flows along the DW.

Let us now consider $p_x+ip_y$ superconducting film in $xy$ plane
so that the crystal anisotropy axis is ${\bf z} \parallel {\bf
c}$. The film is supposed to be thin $d\ll \xi, \lambda$ where
$\xi$ and $\lambda$ are coherence and London penetration lengths.
This condition ensures that we can use the standard approximation
when the magnetic field and order parameter are homogeneous along
the $z$ axis inside the film.

\begin{figure}[!h]
 \centerline{\includegraphics[width=1.0\linewidth]{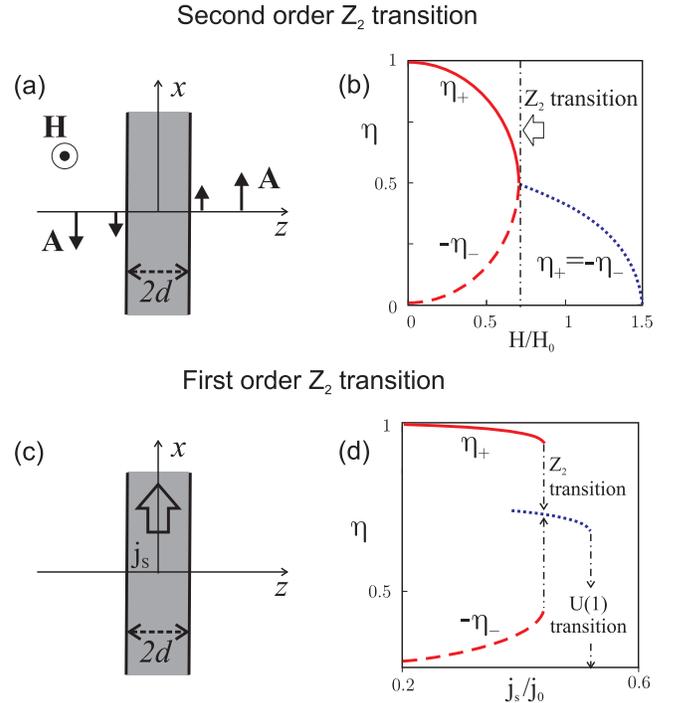}}
 \caption{\label{Fig:Film}
  Phase transitions in a thin film of $p_x+ip_y$ superconductor.
  (a,b) Second order $Z_2$ and $U(1)$ transitions under the action of external magnetic field and
 (c,d) First order transitions in external current.
 By red solid and dashed lines the order parameter amplitudes
 $\eta_+$ and $-\eta_-$ are shown in the $U(1)\times Z_2$ phase.
 The energetically equivalent state is obtained by interchanging
 values of $\eta_+$ and $\eta_-$. The dotted blue line corresponds to the
 non-degenerate $U(1)$ phase with order parameter components $\eta_+=-\eta_+$.
 Magnetic field and current is normalized to
 $H_0 = H_{cm} d / (2 \sqrt{3} \lambda)$ and
  $j_0=(c/4\pi)H_{cm}/\sqrt{2}\lambda$ correspondingly. }
 \end{figure}

First we assume that the film is subjected to the magnetic field
parallel to the film plane ${\bf H} = H {\bf y}$ as shown in
Fig.\ref{Fig:Film}(a). In a thin film of conventional
superconductor the $U(1)$ symmetry braking phase transition  is
known to be of the second order and the critical field $H_c=
\sqrt{6}H_{cm} \lambda/d$ \cite{SchmidtBook}. However in
$U(1)\times Z_2$ superconductor one can expect qualitatively new
features. Indeed the in-plane Meissner current couples the
$L_{z}=\pm 1$ order parameter components. Thus at a certain
critical field $H=H_{Z_2}$ the coupling can be so strong to remove
the $Z_2$ degeneracy of superconducting state. Such symmetry
restoration occurs via the second-order phase transition which is
determined by the coherence length $\xi_{Z_2}$ which is naturally
connected with the size of DW between different chiral domains. At
the point of $Z_2$ phase transition the DW width $\xi_{Z_2}$
diverges and chiral domains disappear. The hierarchical models of
second order phase transitions with sequential breaking of
multiple symmetries were discussed a lot in application to
superconducting heavy fermion compounds
\cite{HeavyFermion,Joynt,Sauls,Machida,JoyntMineev}. Here we
consider another possibility to drive multiple transitions with
external magnetic field and will focus on the physics of non
equilibrium $Z_2$ phase transition.

The proposed scenario can indeed be confirmed by straightforward
calculation. At first we consider an auxiliary problem. Suppose
the Cooper pairs have constant velocity directed along $x$-axis.
Then the order parameters can be represented as
 $ \eta_\pm = \psi_\pm e^{i \kappa k x}$
 where $k$ is a dimensionless Cooper pair velocity. Minimizing the
free energy~(\ref{eq:free_energy_density}) by the amplitudes
$\psi_\pm$ at fixed $k$ we obtain two stable  branches of the
order parameter.

  {\bf (i)} On the first branch the magnitude of order parameter components is different
  $|\psi_+| \neq |\psi_-|$ and they have opposite signs
        \begin{equation}
            \label{eq:solution1}
            \left|{\psi_\pm}\right|^2 = \frac{1}{2}\left [1 - k^2 \pm \sqrt{\left(1 - k^2\right)^2 - k^4 \left (
            \frac{1 + \nu_2}{1 + \nu_1}\right)^2} \right ]\\
        \end{equation}
        Due to the invariance of GL theory (\ref{eq:free_energy_density}) with respect to the replacement of $\psi_+$ to $\psi_-$,
        and vice versa, the found solution is twice degenerate and corresponds to the superconducting  $U(1)\times Z_2$ phase.
        This solution is stable if the velocity of Cooper pairs smaller than the critical value
        $k<k_{Z_2} = \sqrt{(1 + \nu_1)/(2 + \nu_1 + \nu_2)}$.
        Note that $k_{Z_2}<k_{c} = \sqrt{2/(1 - \nu_2)}$
        where $k_c$ is the deparing superfluid velocity which
        destroys superconducting state completely.

 {\bf (ii)} On the second branch the magnitudes of order parameter components
 are the same  $\psi_+ =-\psi_-$  where
        \begin{equation}
            \label{eq:solution2}
            \left|{\psi_\pm}\right|^2 = [1 - k^2 (1 - \nu_2)/2]/(3 +
            \nu_1)
        \end{equation}
        Unlike the previous case, this solution is nondegenerate. Therefore it corresponds to usual $U(1)$
        superconducting state. This phase is stable in the
        interval $k_{Z_2} < |k| < k_{c}$.

 That is we obtain an additional phase transition at
$k=k_{Z_2}$ when the ground state double degeneracy is removed and
the corresponding discrete $Z_2$ symmetry is restored. The order
parameter components change continuously while we shift the $k$
value through the $Z_2$ critical point therefore this is a second
order phase transition.

The solution of an auxiliary problem considered above can be
 applied to find the critical fields of a thin
$p_x+ip_y$ superconducting film. Indeed we choose Landau gauge
$A_x = B_y z$ [see Fig.\ref{Fig:Film}(a)] and use a standard thin
film approximation assuming $\eta_\pm$ to be constants with
respect to $z$ coordinate. Taking the $z$ average of the free
energy yields an effective superfluid velocity $k = \sqrt{\langle
A_x^2 \rangle }  = d H / \sqrt{6}$. Then one immediately find the
critical fields values:
\begin{eqnarray}
    \label{eq:critical_fieldZ2}
        H_{Z_2} = (2\sqrt{3} \lambda /d) k_{Z_2} H_{cm}\\
        \label{eq:critical_fieldU1}
        H_{c}   = (2\sqrt{3} \lambda /d) k_{c}   H_{cm}
\end{eqnarray}

The critical field $H_{Z_2}$ (\ref{eq:critical_fieldZ2}) restores
discrete $Z_2$ symmetry and the field (\ref{eq:critical_fieldU1})
is a standard critical field of thin superconducting film which
suppresses superconductivity completely. The evolution of order
parameter components as functions of applied magnetic field is
shown in Fig.\ref{Fig:Film}(b). In this case both $Z_2$ and $U(1)$
phase transitions are of the second order and characterized by
vanishing order parameters and divergent coherence lengths.

Naturally the order parameter of $Z_2$ phase transition can be
chosen in the form $\eta_1 = \left(\eta_+ + \eta_-\right) / 2$.
Indeed $\eta_1$ vanishes near $H_{Z_2}$ in the first phase and is
identical zero in the second phase. To reveal the physical origin
of $Z_2$ coherence length let us consider the structure of DW in
the vicinity of the critical point. Here we can derive the
equation for the order parameter $\eta_1$
 taking the other component $\eta_2 = \left(\eta_+ - \eta_-\right) /
 2$ to be constant $\eta_2=\eta_2 (H=H_{Z_2})$.
 In this way we assume the order parameter amplitude to be slowly varying real valued function
 $\eta_1=\eta_1(x,y)$ and obtain single component GL equation:
\begin{equation}\label{Eq:GLZ2}
 -D\nabla_r^2 \eta_1+ a \eta_1/2+ b \eta_1^3 = 0
\end{equation}
with coefficients $D=(3+\nu_2)\kappa^{-2}$,
$a=(1+\nu_2/3)(H^2-H^2_{Z_2})d^2$ and $b=2(3+\nu_1)$. We can find
a DW structure as the topological soliton in Eq. (\ref{Eq:GLZ2})
 $ \eta_1 = \sqrt{a/b} \tanh\left(\sqrt{a/D} x\right)$.
Since $a\sim(H_{Z_2}-H)$ we see that the DW dissolves near the
critical field $H_{Z_2}$ and the size of DW proportional to
$\xi_{Z2}\sim(H_{Z_2}-H)^{-1/2}$.

The obtained $Z_2$ symmetry breaking phase transition provides a
unique possibility to create arbitrary concentration of DW in
$p_x+ip_y$ superconductor.
 We  employ a generalization of Kibble-Zurek defect formation
 mechanism\cite{Kibble,Zurek} to explore the DW appearance during
 non-equilibrium $Z_2$ symmetry breaking phase transition.
Let us assume that the external field decreases with the constant
rate $\tau_{H}$ so that $H (t)= (1-t / \tau_{H})H_{Z_2}$. Just
below the $Z_2$ critical point $H<H_{Z_2}$ the growth of $Z_2$
order parameter fluctuations can be described by linearized TDGL
equation \cite{VolovikTDGL,KopninThuneberg}
 \begin{equation}
    \label{eq:TDGL_eta1_real}
    \tau  \eta^\prime_{1t}=\left[H_{Z_2}^2 -H^2(t)\right]\eta_1 +
    6(d\kappa)^{-2}\nabla^2_r \eta_1
\end{equation}
 Eq.(\ref{eq:TDGL_eta1_real}) describes two competing effects:
exponential growth and diffusive spreading due to the last term in
the r.h.s. Comparing these times we can obtain the distance
between defects just after the phase transition as the minimal
length scale which can grow. The characteristic growth time is
$t_Z \sim \sqrt{\tau \tau_H}/H_{Z_2}$, also known as Zurek
time\cite{Zurek,VolovikTDGL}.
  This time should be much less
 than diffusive time $(\kappa dl)^2 \tau$, where $l$ is characteristic length scale. So we obtain the condition
 on the distance between defects immediately after the system has been driven through $Z_2$ phase transition
 $l \sim \left(\tau_H/\tau\right)^{1/4}$.
Thus varying the rate $\tau_H$ it is possible to create arbitrary
concentration of DWs.

Applying an external transport current ${\bf j_s}$ along the film
plane [see Fig.\ref{Fig:Film}(c)] it is possible to obtain the
first order $Z_2$ symmetry breaking phase transition.
 To study this case we introduce a new thermodynamic potential performing
Legendre transformation to the free energy
 $\tilde{f}= f- {\bf k j_s} $ where ${\bf k}$ is
 dimensionless superfluid velocity.
In this case stable state can be found only numerically. An
example of resulting stable branches is shown in
Fig.\ref{Fig:Film}(d) where we plot order parameter components as
functions of the superconducting current density. By red solid and
dashed lines we show order parameter components $\eta_+$ and
$-\eta_-$ for $Z_2$ symmetry breaking branch. By blue dotted line
the non-degenerate state with $\eta_+=-\eta_-$ in $Z_2$ symmetric
phase is shown.

From Fig.\ref{Fig:Film}(d) one can see that $Z_2$ transition is of
the first order so that $U(1)\times Z_2$ and $U(1)$ phases can
coexist. At the same time it is well known that $U(1)$ phase
transition in thin superconducting film with external current is
also of the first order\cite{SchmidtBook}. Thus to have an
additional $Z_2$ symmetry braking phase transition the critical
current of $U(1)\times Z_2$ state should be smaller than that of
$U(1)$. Otherwise the system will fall into normal phase directly
from $U(1)\times Z_2$ state. One can obtain that such regime is
realized provided the condition holds
 $ 2 + (1 + \nu_2)^2/(1 + \nu_1) > (3 + \nu_1)^2/(1 - \nu_2)$.
Therefore in a weak coupling model \cite{AgterbergWC} with
$\nu_1=\nu_2$ there is no first order $Z_2$ phase transition in
external current.

The first order $Z_2$ phase transition discussed above occurs
through the growth of the nuclei with sizes larger than the
critical one\cite{FirstOrderKinetic}. It can be easily estimated
as $\tilde{f}_{s} / \Delta \tilde{f}_{b}$, where $\tilde{f}_{s}$
is the surface free energy density and $\Delta \tilde{f}_{b}$ is
difference of bulk free energy densities in two phases. Thus the
critical size is determined by the external current through the
bulk energy dependence
 $\Delta \tilde{f}_{b}=\Delta \tilde{f}_{b} (j_s)$. It is natural to expect that the
distance between DW after the first order transition should be
determined by the critical size which can vary from $0$ to
$\infty$ by setting the current $j_s$.

Finally let us discuss a way to measure the residual DW
concentration which survives after the transient processes after
the non-equilibrium $Z_2$ phase transition. The DW can be
stabilized by geometrical confinement in mesoscopic
samples\cite{Vakarouk}, pinning on vortices and defects
\cite{Ichioka,Golbardt,SigristAgterbDW}.
 Besides several known experimental approaches \cite{Kallin,Josephson,Kerr}
 we suggest to employ
 transport measurements in the mixed state produced by magnetic field ${\bf H} \parallel {\bf c}$ where ${\bf c}$ is anisotropy axis.
  The proposed method is based on the observation that
 such field creates Abrikosov vortices which are known to remove $Z_2$ degeneracy of superconducting vacuum in $p_x+ip_y$
 superconductor. That is vortices have different core structures in the chiral domains with $({\bf H L})>
 (<)0$ \cite{MelnikovBarash,AgterbergVortices,SigristVortices}
 where ${\bf L}$ denotes the direction of the internal orbital momentum of Cooper pairs which in
 our case is ${\bf L} \parallel {\bf c}$. We denote these
 vortex structures $N_{+}$ and $N_{-}$ vortices
 correspondingly.

In isotropic case $\nu_1=\nu_2=0$ the order parameter in axially
symmetric vortices has form $\eta_\pm = \left|{\eta_\pm}\right|
(r)e^{im_\pm \theta}$ where $(r, \theta)$ are polar coordinates
with the origin at the vortex center. Axial symmetry is preserved
provided the choice of the vorticityies $m_+=1$, $m_-=3$ for $N_+$
and $m_+=1$, $m_-=-1$ for $N_-$ vortices.

 Here we note that $N_+$
and $N_-$ vortices have different vi1scosities
 due to the difference in their core structures.
Hence the flux flow conductivity has a chirality sensitive
contribution $\sigma=\sigma_0+\sigma_1 ({\bf HL})$. The flux fow
conductivity can be calculated within the framework of time
dependent GL theory \cite{KopninBook}. In this way we obtain
    \begin{equation}
    \label{eq:conductivity_ohm}
    \sigma/\tilde{\sigma} =
    \int\limits_0^\infty \sum\limits_{\alpha} \left [
    \rho  \left|{\eta_\alpha}\right|^{\prime 2}_\rho  + \left|{\eta_\alpha}\right|^2 \left (m_\alpha^2 + \rho\mu_0 \right)
    \right ]\; d\rho
    \end{equation}
Here we normalize conductivity by $\tilde{\sigma}=\sigma_n u
\kappa H_{cm}/\sqrt{2} H$, where  $l$ is the electric field
penetration length $l^2 = (\sigma_n \Phi_0^2)/8 \pi^2 c^2 \tau$,
$u = (\xi/l)^2$,  $\rho=r/l$ and $\sigma_n$ is a normal metal
conductivity. The function $\mu_0=\mu_0(r)$ determines
electrostatic potential around moving vortex $ \varphi = \mu_0(r)
({\bf e_r}{\bf \left[v, z_0\right]}$) where ${\bf v}$ is vortex
velocity and ${\bf e_r}={\bf r}/r$. It satisfies the Poisson
equation
    \begin{equation}
    \label{eq:potential_equation}
    \left( \nabla^2_\rho    -
    \rho^{-2} - \left|{\eta_+}\right|^2 - \left|{\eta_-}\right|^2 \right ) \mu_0=
    \rho^{-1}m_\alpha \left|{\eta_\alpha}\right|^2
    \end{equation}
 For example taking the parameters
 $\kappa = 2.3$ and $u = 6$ we obtain the flux-flow conductivities
 $\sigma_+ = 13.5\sigma_n H_{cm}/H$ and $\sigma_- =14.6 \sigma_n H_{cm}/H$ for
$N_+$ and $N_-$ vortices correspondingly so that the chirality
sensitive part is $\sigma_1=(\sigma_+-\sigma_-)/2=0.018 \sigma_0$.
 Averaged over the sample flux flow conductivity is given by
$\bar{\sigma}= \sigma_+S_+ + \sigma_-S_-$ where $S_\pm$ are the
measures of the parts occupied by domains of positive and negative
chiralities. Thus measuring flux flow conductivity $\sigma$ it is
possible to study the evolution of domain structure in $Sr_2RuO_4$
generated through the nonequilibrium $Z_2$ phase transition.

To conclude we have found discrete symmetry breaking phase
transition in $p_x+ip_y$ superconductors. The transition can be of
the first order if driven by external current and of the second
order under the action of external field. That is applying
in-plane magnetic field to the thin superconducting film one can
drive it continuously from $U(1)\times Z_2$ to the simple $U(1)$
state. Such $Z_2$ symmetry restoration is marked by dissolution of
DWs. Decreasing the field through $Z_2$ critical point at a
constant rate one can create a particular concentration of DWs
according to the Kibble-Zurek scenario. This possibility can
facilitate experimental identification of DWs. Results on the
present paper have been derived for a thin superconducting film.
Our approach can be generalized to describe surface layers with
thickness of the order of London penetration length in
superconducting single crystals.

We thank prof. Alexander Mel'nikov for many stimulating
discussions. This work was supported by Russian Foundation for
Basic Research Grants No 11-02-00891, 13-02-97126. MS was
supported by Russian President scholarship (SP- 6811.2013.5).

\end{document}